\documentclass[prd, aps, superscriptaddress, preprintnumbers, twocolumn, floatfix, nofootinbib]{revtex4}

\usepackage{amsfonts}
\usepackage{amsmath}
\usepackage{amssymb}
\usepackage{bm}
\usepackage{dcolumn}
\usepackage{graphicx}   
\usepackage[latin1]{inputenc}
\usepackage{latexsym}
\usepackage{rotating}
\usepackage{hyperref}
\usepackage{graphicx}
\usepackage{color}

\usepackage{amsfonts}
\usepackage{amsmath}
\usepackage{amssymb}
\usepackage{bm}
\usepackage{dcolumn}
\usepackage{graphicx}
\usepackage[latin1]{inputenc}
\usepackage{latexsym}
\usepackage{rotating}
\usepackage{hyperref}

\newcommand\be{\begin{equation}}
\newcommand\ba{\begin{eqnarray}}
\newcommand\ee{\end{equation}}
\newcommand\ea{\end{eqnarray}}

\begin{document}

\title {Constraints on Superconducting Cosmic Strings from the Global $21$-cm Signal before Reionization}

\author{Robert Brandenberger}
\email{rhb@physics.mcgill.ca}
\affiliation{Physics Department, McGill University, Montreal, QC, H3A 2T8, Canada}

\author{Bryce Cyr}
\email{bryce.cyr@mail.mcgill.ca}
\affiliation{Physics Department, McGill University, Montreal, QC, H3A 2T8, Canada}

\author{Rui Shi}
\email{ustcwysr@mail.ustc.edu.cn}
\affiliation{School of Physical Sciences, University of Science and Technology of China, Hefei 230026, China\\
and Physics Department, McGill University, Montreal, QC, H3A 2T8, Canada}

\date{\today}

\begin{abstract}

Electromagnetic radiation from the cusp region of superconducting cosmic strings leads to a radio excess in the photon spectrum in the early universe and can produce a deep absorption feature in the global 21cm signal before the epoch of reionization. We study the constraints on the parameter space of superconducting strings which can be derived by demanding that the absorption feature is not larger in amplitude than what has recently been reported by the EDGES collaboration.

\end{abstract}

\maketitle

\section{Introduction} 

There has recently been a lot of work exploring the observational signals of cosmic strings (see e.g. \cite{new} for a review of recent work). Cosmic strings are one-dimensional topological defects which are solutions of the field equations in many theories beyond the Standard Model of particle physics (see e.g. \cite{CSrevs} for reviews of the physics and cosmology of cosmic strings). If Nature is described by a theory which admits cosmic string solutions, then a network of cosmic strings will form during a phase transition in the early universe and persist until the present time \cite{Kibble}. Strings lead to interesting signatures for a variety of cosmological observations. Not observing these signals will lead to constraints on the parameter space of particle physics models beyond the Standard Model \cite{BSM}.

Most of the recent studies of the cosmological imprints of cosmic strings have been done in the context of strings which only interact gravitationally. In this case, the network of strings is described by a single parameter, the string tension $\mu$ (usually expressed in terms of the dimensionless number $G\mu$, where $G$ is Newton's gravitational constant). However, in many models, strings carry conserved electromagnetic currents, either bosonic or fermionic, and they are hence superconducting \cite{Witten}. In a theory which has superconducting strings, currents on the string are induced during the phase transition which produces the strings, and during the later evolution of the strings through the plasma of the early universe. Superconducting cosmic strings emit electromagnetic radiation. This radiation is dominated by the cusp region on cosmic string loops, regions which move with a velocity which approaches the speed of light. Cusps are generic features of string loops, with  at least one cusp per oscillation period forming on a string loop \cite{KT}. The bursts of electromagnetic radiation from strings cusps were studied e.g. in \cite{AVTV}, \cite{Spergel}, \cite{Copeland} and \cite{Olum1}. The fact that this emission gives rise to spectral distortions of the CMB was investigated in \cite{Sanchez} and \cite{Tashiro1}. The connection with gamma ray bursts was explored in \cite{GRB}, and the connection with fast radio transients in \cite{FRB} \footnote{For earlier work on the connection between superconducting strings and ultra-high energy cosmic rays see \cite{UHECR}.}.

Causality implies \cite{Kibble} that at all times after the phase transition during which the strings are formed, there will be at least one ``long'' string crossing each Hubble patch of space. Here, ``long'' refers to a string with curvature radius larger than the Hubble radius. Analytical arguments \cite{CSrevs} indicate that the network of long strings approaches a ``scaling'' solution, according to which the distribution of strings is statistically the same at all times if lengths are scaled to the Hubble radius. This is confirmed by numerical simulations of the evolution of cosmic string networks \cite{CSsimuls}. The network of long strings maintains its scaling property by losing energy to string loops during long string intersections. Hence, a distribution of string loops with a large range of radii $R$ builds up. Analytical arguments (this time not supported by general causality considerations) indicate that the distribution of string loops will also approach a scaling solution. At times later than $t_{eq}$, the time of equal matter and radiation, the distribution of string loops is given by (see e.g. \cite{CSloop})
\be
n(R, t) \, = \, \nu R^{-5/2} t_{eq}^{1/2} t^{-2} \,\,\, {\rm{for}} \,\,\, R_c(t) \, < \, R \, < \alpha t_{eq} \, ,
\label{Rdist}
\ee
and
\be
n(R,t) \, \sim \, R^{-2} t^{-2} \,\,\, {\rm{for}} \,\,\, R \, > \, \alpha t_{eq} \, ,
\ee
where $\nu$ is a constant which depends on the number of long string segments crossing a Hubble volume, $R_c(t)$ is a cutoff radius below which loops live less than one Hubble expansion time, and $\alpha$ is a constant indicating at what fraction of the Hubble radius loops are produced.

For non-superconducting strings, the dominant energy loss mechanism of string loops is gravitational radiation. Since string loops have relativistic tension, they oscillate and emit gravitational radiation. The power of gravitational radiation from a string loop is \cite{grav}
\be
P_{gw} \, = \, \gamma G \mu^2 \, ,
\ee
where $\gamma$ is a constant which, according to numerical simulations of gravitational wave emission \cite{grav} is of the order $\gamma \sim 10^2$.

It can be shown that the total power of electromagnetic radiation from cusp regions, averaged over an oscillation period, is \cite{TC}
\be
P_{em} \, = \, \kappa I \sqrt{\mu} \, ,
\ee
where $I$ is the current on the string and $\kappa$ is a constant of the order $1$. For a fixed value of $G\mu$ there is a critical current $I_c$ above which electromagnetic radiation dominates, and below which gravitational radiation is more important. This critical current is given by
\be \label{critcur}
I_c \, = \, \kappa^{-1} \gamma (G\mu)^{3/2} m_{pl} \, ,
\ee
where $m_{pl}$ is the Planck mass. If gravitational radiation dominates, then the cutoff radius $R_c(t)$ is given by
\be \label{RcGW}
R_c(t) \, = \, \gamma \beta^{-1} G\mu t \, ,
\ee
where $\beta$ is a constant which gives the mean loop length in terms of the radius (for a circular loop we would have $\beta = 2\pi$), while if electromagnetic radiation dominates we have
\be \label{RcEM}
R_c(t) \, = \, \kappa \beta^{-1} I \mu^{-1/2} t \, .
\ee

There is another energy loss mechanism for string loops: cusp decay \cite{decay}. This mechanism is more important than gravitational radiation loss only for very low tension strings, and more important than electromagnetic radiation only for very small currents. Thus, the mechanism is only important in regions of parameter space where the electromagnetic radiation is small. Hence, in this work we will not consider the cusp decay channel.

The electromagnetic emission is non-thermal and concentrated at low frequencies. We consider the enhanced energy density in electromagnetic radiation at frequencies at and below the frequency of 21cm radiation. This radio excess leads to a deep absorption signal in the global 21cm signal before reionization. Such a signal was recently reported by the EDGES experiment \cite{Edges} with an absorption deeper than predicted by the standard cosmological paradigm. A radio excess could possibly explain this signal. In an earlier paper \cite{us}, we studied the contribution of electromagnetic radiation from cusp decays of non-superconducting strings to the radio excess and found it to be much too small to effect observations. Superconducting strings, on the other hand, produce much more electromagnetic radiation. Here we study the contribution to a possible radio excess which these strings can produce. By demanding that the global absorption is smaller than what has been measured by the EDGES collaboration \cite{Edges} we can derive constraints on the parameter space of superconducting strings. We find a region of parameter space which is indeed ruled out by observations.

In the following we will first review how a soft radio photon excess can lead to a large absorption feature in the 21cm global signal. In Section 3, we then compute the radio excess predicted by a network of superconducting cosmic string loops as a function of the cosmic string parameters $G\mu$ and $I$, determine their fractional contribution to the total radio spectrum of the cosmic microwave background (CMB), and derive the constraints on the parameter space which results from demanding that the induced 21cm absorption signal is not larger than what has been detected. We discuss our results and compare with previous studies of cosmological constraints on superconducting cosmic strings in the final section.

We work in natural units in which the speed of light, Planck's constant and Boltzmann's constant are set to $1$. We work in the context of standard cosmology. The variable $t$ is used for physical time, $t_{eq}$ denoting the time of equal matter and radiation, and $t_{rec}$ the time of recombination (which is later than the time of equal matter and radiation). The temperature of the radiation gas is denoted by $T$.

\section{Radio Excess and Global $21$cm Absorption Signal}

One way in which the parameter space of superconducting cosmic strings can be constrained is by the amplitude of the absorption feature in the global 21cm signal. Observing the universe through the 21cm window can provide information about the distribution of neutral hydrogen at early times, in particular before and during reionization. CMB photons passing through a cold cloud of neutral hydrogen are absorbed by exciting the hydrogen hyperfine transition (see e.g. \cite{Furl} for an in-depth review of the physics of 21cm surveys). In an early paper \cite{Oscar} it was pointed out that the global 21cm signal can provide a tool to probe for the existence of cosmic string wakes before reionization.

There has recently been a lot of interest in the global 21cm absorption signal before the epoch of reionization as a consequence of the detection of an unexpectedly large absorption signal by the EDGES collaboration \cite{Edges}. The detected amplitude of the temperature decrement is at least twice what is expected in standard cosmology. Assuming a mostly neutral IGM, the temperature decrement $\delta T_b$ is given by the following proportionality
\begin{align}
\delta T_b \, \propto \, 1-\frac{T_{\gamma}}{T_{spin}} \, ,
\end{align}
where $T_{\gamma}$ is the effective temperature of the photons at the 21cm frequency, and $T_{spin}$ is the spin temperature of the hydrogen gas. The effective temperature of the photons, in turn, is given by
\be
T_{\gamma} \, = \, T_{CMB} + T_{21} \, ,
\ee
where $T_{CMB}$ is the overall CMB black body temperature, and $T_{21}$ is the effective temperature due to extra photons at 21cm frequencies. These effective temperatures are given in terms of the energy densities of photons in the frequency range of 21cm. Given a new source of radio photons, the boost in the amplitude of the absorption feature is given by \cite{Edges2} the factor
\be
{\cal{F}} \, \sim \, 1 + \frac{T_{21}}{T_{CMB}} \, .
\ee

Following the results reported in \cite{Edges} there has been a flurry of work proposing ways to explain the extra absorption. A lot of the proposed models have focused on ways to reduce the spin temperature, e.g. by invoking dark matter interactions (see e.g. \cite{DMearly} for early work). On the other hand, it was pointed out in \cite{Holder} that extra 21cm absorption can also be obtained if there is a new source of radio photons. In fact, an excess of the radio background has been reported by the ARCADE-2 experiment \cite{Arcade}. This serves as an additional piece of motivation, as the existence of cosmic strings (both superconducting and not) would produce an anomalous background.

In this work we will compute the contribution of superconducting cosmic string loops to the radio photon background, and we will provide bounds on the parameter space of such strings by demanding that the absorption feature produced by the strings not exceed the amplitude reported in \cite{Edges} \footnote{In a recent paper \cite{us} we studied the radio excess produced by the cusp decay of non-superconducting strings, but we found that the predicted radio spectrum is too low in amplitude to yield constraints on the cosmic string parameter space.}. Note that if the global $21{\rm{cm}}$ signal turns out to be lower than what was reported, our bounds will become stronger.

\section{Radio Excess from Superconducting Cosmic Strings}

In the following we will compute the energy density in radio photons due to the emission from a network of loops of superconducting cosmic strings. The energy density in photons in the frequency interval $[0, \omega_{21}]$ at time $t$ is
\be
\rho^{21}(t) \, = \, \int_{t_{rec}}^t d\tau \frac{d\rho^{21}(\tau)}{d\tau} \bigl( \frac{\tau}{t} \bigr)^{8/3} \, ,
\ee
where the first term inside the integral is the energy density in photons inserted per unit time, and the second factor gives the cosmological redshift of the radiation energy density in the matter dominated epoch. The energy density per unit time is given by integrating over all frequencies below the frequency $\omega_{21}$ blueshifted to time $\tau < t$
\be \label{int1}
\omega_{21}(\tau) \, = \, \omega_{21} \bigl( \frac{t}{\tau} \bigr)^{2/3}
\ee
and integrating over the distribution of string loops. This gives
\be \label{int2}
\frac{d\rho^{21}(\tau)}{d\tau}\, = \, \int_0^{\omega_{21}(\tau)} d\omega 
\int_{R_c(\tau)}^{\alpha\tau}dR n(R, \tau) \frac{dP}{d\omega}(\omega) \, ,
\ee
where the final factor is the power per unit frequency from cosmic string core region emission, and is given by \cite{TC}
\be
\frac{dP}{d\omega}(\omega) \, = \, \kappa I^2 R^{1/3} \omega^{-2/3} \, .
\ee

The integral over $R$ in (\ref{int2}) is dominated by the smallest value of $R$ greater that the cutoff value $R_c(t)$. Let us focus on values of $G\mu$ and $I$ which are sufficiently small such that the value of $R_c(t)$ at the time of reionization is smaller than $\alpha t_{eq}$. In this case, we can use the expression (\ref{Rdist}) for $n(R, \tau)$. Inserting this and performing the integral over $R$ we obtain
\be
\rho^{21}(t) \, = \, 3 {\tilde{\kappa}} \nu I^2 \omega_{21}^{1/3} t_{eq}^{1/2} t^{-2}
\int_{t_{rec}}^t d\tau R_c(\tau)^{-7/6} \bigl( \frac{\tau}{t} \bigr)^{4/9} \, ,
\ee
where ${\tilde{\kappa}}$ is $\kappa$ multiplied by constants of order 1.

For currents smaller than the critical current $I_c$ (which is a function of $G\mu$), we insert the value of the cutoff radius given by (\ref{RcGW}) and obtain
\be \label{result1}
\rho^{21}(t) \, = \, 18 {\tilde{\kappa}} \nu I^2 \omega_{21}^{1/3} t_{eq}^{1/2} t^{-2}
\gamma^{-7/6} \beta^{7/6} (G\mu)^{-7/6} t^{-1/6}\, .
\ee
Inserting the maximal current for which this expression is valid, namely (\ref{critcur}) yields
\be \label{result2}
\rho^{21}(t) \, = \, 18 {\tilde{\kappa}} \kappa^{-2} \nu G^{-1} \omega_{21}^{1/3} t_{eq}^{1/2} t^{-2}
\gamma^{5/6} \beta^{7/6} (G\mu)^{11/6} t^{-1/6}\, .
\ee

For currents larger than the critical one we insert the value of the cutoff radius given by (\ref{RcEM}) and obtain
\be \label{result3}
\rho^{21}(t) \, = \, 18 {\tilde{\kappa}} \kappa^{-7/6} \nu I^{5/6} \omega_{21}^{1/3} t_{eq}^{1/2} t^{-2}
\beta^{7/6} (G\mu)^{7/12} G^{-7/12} t^{-1/6}\, .
\ee
Inserting the current (\ref{critcur}), the minimal current for which this computation applies, yields the same result we obtained above, namely (\ref{result2}).

We now compare the magnitude of the above radio excess of photons at frequencies smaller than $\nu_{21}$ with the CMB black body photons in this frequency range. Since the value $\omega_{21}$ is in the low frequency (Rayleigh-Jeans) regime of the black body for the temperature at the time of reionization, the energy density in CMB photons is given by
\be
\rho_{CMB}(t) \, = \, \frac{1}{3\pi^2} \omega_{21}^3 T(t) \, .
\ee
To compute the ratio ${\cal{R}}$ of $\rho^{21}(t)$ and $\rho_{CMB}(t)$ we make use of the Friedmann equation to relate time and temperature. Specifically, we use
\be
G^{-1} t^{-2} \, = \, 6 \pi \bigl( \frac{t}{t_{eq}} \bigr)^{2/3} T^4 \, .
\ee
Inserting the value of $I_c$ into $\rho^{21}(t)$ yields
\be
{\cal{R}} (I = I_c) \, = \, {\cal{A}} \nu \gamma^{5/6} \beta^{7/6} (G\mu)^{11/6} 
\bigl( \frac{t}{t_{eq}} \bigr)^{1/18} \bigl( \frac{T}{\omega_{21}} \bigr)^{8/3} \bigl( \frac{m_{pl}}{T} \bigr)^{1/3} \, ,
\ee
where ${\cal{A}}$ is a constant of the order $10^3$ (assuming that $\kappa$ is a constant of the order 1), and where we have left the dependence on $\nu, \beta$ and $\gamma$ explicit. For typical values $\beta = \gamma = \nu = 10$, and evaluating the result at the temperature of reionization we obtain
\be \label{ratio1}
{\cal{R}} (I = I_c) \, \sim \,  10^{12} (G\mu)_6^{11/6}
\ee
where $(G\mu)_6$ is the value of $G\mu$ in units of $10^{-6}$.

For values of the current larger than $I_c$ we have
\be \label{ratio2}
{\cal{R}}(I) \, = \, \bigl( \frac{I}{I_c} \bigr)^{5/6} {\cal{R}}(I_c) \,\,\, (I > I_c) \, ,
\ee
while for values smaller than $I_c$ we have
\be \label{ratio3}
{\cal{R}}(I) \, = \, \bigl( \frac{I}{I_c} \bigr)^{2} {\cal{R}}(I_c) \,\,\, (I < I_c) \, .
\ee

If the EDGES results are to be explained by excess photons at low frequencies, then the value of ${\cal{R}}$ should be of the order $1$ \cite{Edges2}. If we demand that the distribution of superconducting string loops does not produce an absorption feature larger in amplitude than has been observed, the parameter space of strings must be constrained to yield ${\cal{R}} \leq 1$. From (\ref{ratio1}), (\ref{ratio2}) and (\ref{ratio3}) we see that a significant parameter space of high tension and high current superconducting cosmic strings is ruled out.

Our results are summarized in Figure 1. In this figure, the vertical axis is the value of $G\mu$ and the horizontal axis the current $I$. The shaded region bounded by the solid line is the region of parameter space which is ruled out by the current data. The dashed line indicates the boundary of the parameter space which would be ruled out if the absorption signal were to correspond to a value of ${\cal{R}} = 0.1$ instead of ${\cal{R}} = 1$. The diagonal line corresponds to the critical current $I_c$. In the region above this line the decay of string loops is dominated by gravitational radiation, in the region below it is dominated by electromagnetic radiation.

\begin{figure*}[ht]
\includegraphics[width=\linewidth]{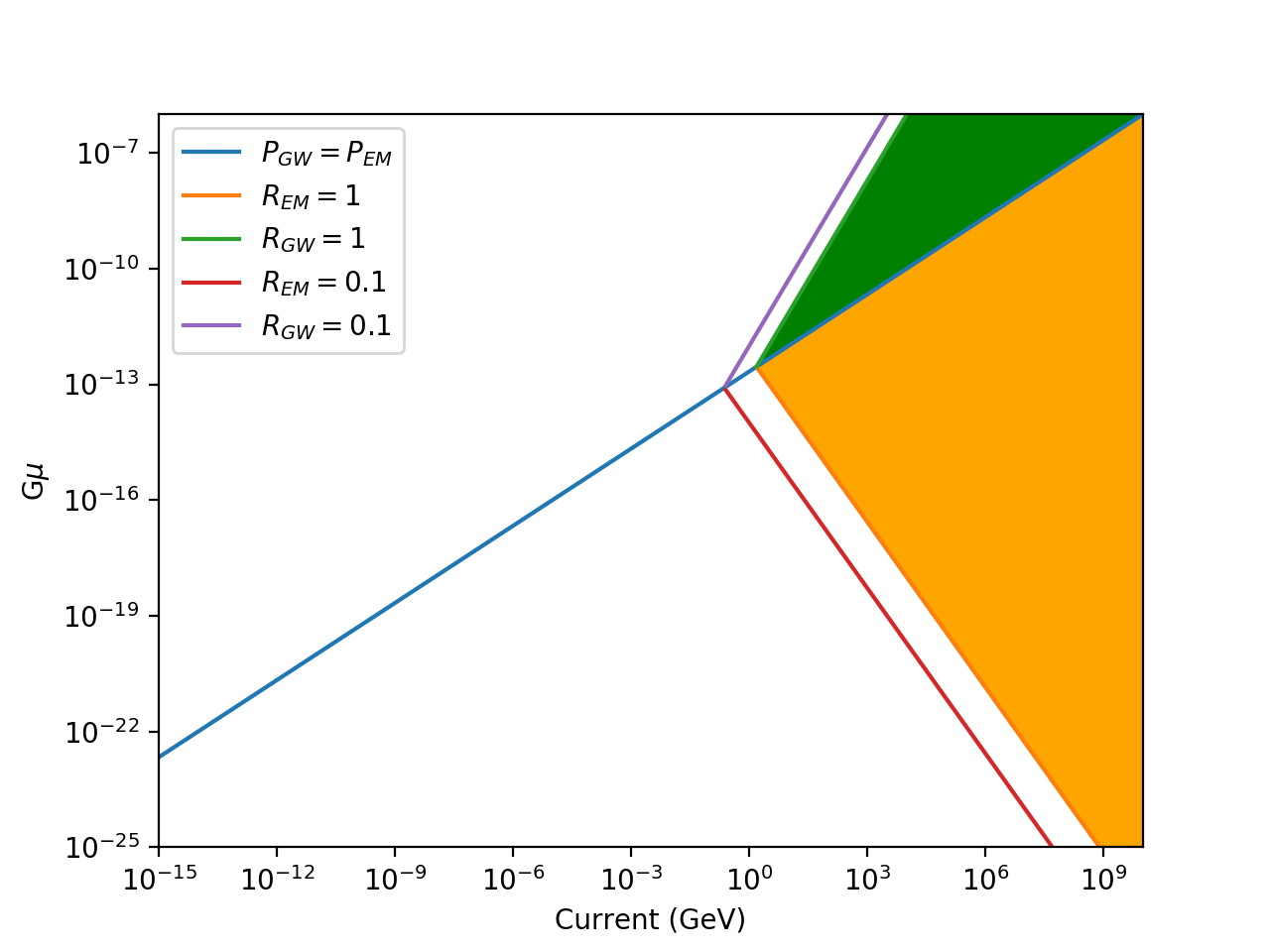}
\caption{The shaded regions show the parameter
space of superconducting cosmic strings which are ruled out
by our analysis, assuming that the global 21cm absorption signal
is not larger than what has been reported by the EDGES collaboration.
The regions are bounded by the curves along which ${\cal{R}} = 1$
(see the main text for the definition of ${\cal{R}}$).
If the signal is less than $10\%$ of what has been reported, then
the parameter space which is ruled out increases as indicated by
the red and purple lines (which correspond to ${\cal{R}} = 0.1$).
The vertical axis gives the value of $G\mu$,
the horizontal axis is the current. The diagonal blue curve corresponds
to the critical current as a function of $G\mu$. Above the curve
gravitational radiation dominates, below electromagnetic radiation.}
\end{figure*}

Assuming standard values of the constants which arise in the cosmic string distribution
\ba
\gamma \, &=& \, 10 \nonumber \\
\beta \, &=& \, 10 \\
\nu \, &=& \, 10 \nonumber
\ea
we find
\be
I_c \, \sim \, (G\mu)_6^{3/2} 10^{10} {\rm{GeV}} \, .
\ee
For $I = I_c$ the constraint ${\cal{R}} < 1$ implies
\be
(G\mu)_6 \, < \, 10^{-6.5} \, .
\ee
This limit is marginally stronger than the limit on the string tension from pulsar timing array measurements \cite{pulsar}. The pulsar timing limit on $G\mu$, however, assumes that the distribution of string loops is determined by energy loss only from gravitational radiation. Since the pulsar timing constraints are dominated by the smallest loops in the scaling distribution, the bounds are much weaker if $I > I_c$. According to the analysis of \cite{Miyamoto}, the pulsar timing constraints essentially disappear for $I > I_c$. Our analysis lets us set an upper bound on the cosmic string current 
\be
I \, < \, I_c (G\mu)_6^{-66/35} 10^{-72/5} \, \sim \, (G\mu)^{-1/3} 10^{-7} {\rm{GeV}} \, .
\ee

\section{Conclusions and Discussion}

We have derived constraints on the parameter space of superconducting cosmic strings obtained by demanding that the induced absorption feature in the global 21cm signal before reionization not exceed the value observed by the EDGES collaboration \cite{Edges}. Our results are summarized in Figure 1. Note that we have implicitly assumed total efficiency in the Lyman alpha coupling of the 21cm spin temperature to the kinetic temperature of the IGM during the formation of the first stars. If the coupling efficiency were low, then a higher excess radiation would be required to produce a given decrement in the signal, and our bounds on the cosmic string parameter space would weaken. 

For values of the string current $I$ smaller than the critical current $I_c(G\mu)$, our constraints are weaker than the ones which follow from pulsar timing array measurements. The pulsar timing constraints come from gravitational radiation emitted by string loops. The effect is dominated by the smallest loops which survive more than one Hubble time. In the region of parameter space where gravitational radiation dominates over electromagnetic radiation this effective cutoff radius is $\gamma G\mu t$. However, for large current for which electromagnetic radiation dominates the cutoff radius is larger and hence the pulsar timing constraints on $G\mu$ are weaker. The analysis of \cite{Miyamoto} has shown that the pulsar timing constraint essentially disappears for $I > I_c$. In this region, our constraints rule out a large parameter space of strings with large currents.

Before concluding, it is prudent to mention some limitations of our analysis. First, we have remained agnostic to the details of the symmetry breaking scheme that forms these strings. This was done to remain as general as possible, though in doing so we have neglected to describe the best way to generate steady currents on a distribution of loops. The motion of superconducting strings through the universe will naturally produce a range of currents within the loop distribution, as the inhomogeneity of small scale magnetic fields induces different current magnitudes. Future work should consider a distribution of such currents. 

As well, it has been shown that the production of low frequency photons off of cosmic string cusps will accelerate particles in the ISM and IGM to very high Lorentz factors \cite{GRB}. This can cause a hydrodynamical flow of the gas in the vicinity of a string, which is often terminated by a shock. This shock could lead to a gamma ray burst, of which it is unclear how physics at reionization could be effected. This mechanism is very difficult to track analytically, and so a more numerical approach would be beneficial in elucidating the consequences of this effect.

There have been other constraints on superconducting cosmic strings. For example, the electromagnetic radiation from string loops can locally reionize matter and thus change the optical depth which has been measured with great accuracy by the recent CMB experiments (see e.g. \cite{Planck}). The resulting constraints on the parameter space of superconducting strings has been studied in \cite{Tanmay} and \cite{Miyamoto}. In addition, the electromagnetic radiation can lead to CMB spectral distortions, and this provides additional constraints. Comparing our results of Figure 1 with the results of the analysis of \cite{Miyamoto} which summarizes the constraints mentioned above (see in particular the summary figure Figure 8 in \cite{Miyamoto}), we see that our analysis rules out an additional region of parameter space, a region which is particularly interesting for large values of the current.

Non-superconducting cosmic strings can provide nonlinear seeds at high redshifts which could help explain the abundance of super-massive black holes \cite{SMBH}. They could also play a role in the formation of globular clusters \cite{Rebecca}. It would be interesting to study to what extent currents on superconducting strings would change those results.

\section*{Acknowledgements}

The research at McGill is supported in part by funds from NSERC and from the Canada Research Chair program. BC is supported in part by a McGill Space Institute fellowship. RS would like to thank Yifu Cai for helpful discussions. We are grateful to Raul Monsalve for many stimulating discussions and for comments on the manuscript. We also thank Oscar Hernandez, Samuel Lalibert\'e and Disrael Cunha for discussions.

\end{document}